\documentclass[graybox]{svmult}

\usepackage{mathptmx}       % selects Times Roman as basic font
\usepackage{helvet}         % selects Helvetica as sans-serif font
\usepackage{courier}        % selects Courier as typewriter font
\usepackage{type1cm}        % activate if the above 3 fonts are
\usepackage{makeidx}         % allows index generation
\usepackage{graphicx}        % standard LaTeX graphics tool
\usepackage{multicol}        % used for the two-column index
\usepackage[bottom]{footmisc}% places footnotes at page bottom
\newcommand{\be}{\,\begin{equation}}
\newcommand{\ee}{\,\end{equation}}
\newcommand{\apj}{ApJ}
\newcommand{\apjl}{ApJL}
\newcommand{\mnras}{MNRAS}
\newcommand{\aap}{A\&A}

\makeindex             % used for the subject index
\begin{document}

\title*{Positrons from pulsar winds}
\author{Pasquale Blasi and Elena Amato}
\institute{Pasquale Blasi \at INAF/Arcetri Astrophysical Observatory, \email{blasi@arcetri.astro.it}\\
Elena Amato \at INAF/Arcetri Astrophysical Observatory, \email{amato@arcetri.astro.it}
}
\maketitle

\abstract{Pulsars, or more generally rotation powered neutron stars, are excellent factories of antimatter in the Galaxy, in the form of pairs of electrons and positrons. Electrons are initially extracted from the surface of the star by the intense rotation induced electric fields and later transformed into electron-positron pairs through electromagnetic cascading. Observations of Pulsar Wind Nebulae (PWNe) show that cascades in the pulsar magnetosphere must ensure pair multiplicities of order $10^{4}-10^{5}$. These pairs finally end up as part of the relativistic magnetized wind emanating from the pulsar. The wind is slowed down, from its highly relativistic bulk motion, at a termination shock, which represents the reverse shock due to its interaction with the surrounding ejecta of the progenitor supernova. At the (relativistic) termination shock, acceleration of the pairs occurs, as part of the dissipation process, so that the cold wind is transformed into a plasma of relativistic non-thermal particles, plus a potential thermal component, which however has never been observed. As long as the pulsar wind is embedded in the supernova remnant these pairs are forced to escavate a bubble and lose energy adiabatically (because of the expansion) and radiatively (because of magnetic and radiation fields). We discuss here the observational constraints on the energy and number content of such pairs and discuss the scenarios that may allow for the pairs to escape in the interstellar medium and possibly contribute to the positron excess that has recently been detected by the PAMELA satellite. Special attention is dedicated to the case of nebulae surrounding high speed pulsars, observationally classified as Pulsar Bow Shock Nebulae. The pairs produced in these objects may be effectively carried out of the Supernova Remnant and released in the Interstellar Medium. As a result, Bow Shock Pulsar Wind Nebulae might be the main contributors to the positron excess in the Galaxy.} 

\section{Introduction}

Pulsars have long been known to be powerful factories of electron-positron pairs: the rotation of the pulsar leads to an induced electric field that extracts electrons from the star's surface. These electrons lose energy via curvature radiation while propagating far from the star along the magnetic field lines, and the emitted photons are so energetic that an electron-positron pair can be formed in the intense neutron star magnetic field. The process can in principle repeat itself numerous times, increasing the number of pairs up to a large value of the so-called multiplicity, defined as the number of pairs derived on average from a single primary particle that leaves the star surface. How many pairs in the end leave the pulsar magnetosphere is a question that still does not have a definite answer in terms of theory of pulsar magnetospheres, depending on the 
location and structure of the magnetospheric gaps, the regions where there is an unscreened electric field parallel to the local magnetic field and particles can be accelerated. Progress on this topic has recently come from gamma-ray observations, by MAGIC first \cite{magic08} and Fermi \cite{fermicat10} later, which have allowed to locate the source of pulsed high energy emission in the outer magnetosphere, ruling out the possibility that gamma-rays are produced close to the stellar surface (polar cap scenario) together with the low energy radio emission.

As we will discuss below, in spite of the theoretical uncertainties, rather stringent constraints on the pair multiplicity of pulsars can be obtained from observations of their nebulae. The pairs produced in the magnetosphere become part of the relativistic magnetized wind emanating from the pulsar and find themselves in a complex environment. The neutron star was born in a supernova explosion, from the collapse of a massive star, and therefore initially lies in a region bounded by the ejecta of its progenitor and further out by the Supernova blast wave propagating in the Interstellar Medium (ISM). When the cold magnetized wind launched by the star and propagating at almost the speed of light hits the sub-relativistically expanding ejecta, the impact results in a system of shock waves: the outer one propagates in the ejecta, while a reverse shock propagates back towards the star. This is the so-called termination shock, where the wind is slowed down, its bulk energy dissipated and turned into that of a relativistically hot, magnetized fluid, which then shines as a Pulsar Wind Nebula. 

Indeed, the termination shock is also where particle acceleration occurs. When the electron-positron pairs produced in the pulsar magnetosphere reach the termination shock, due to the severe energy losses they have suffered, their energy is not larger than few tens of GeV, in spite of the enormous potential drop available, which might be as large as 10$^{16}$ V. At the termination shock, a relatively large fraction (up to few tens of percent) of the wind bulk energy is converted into accelerated pairs, which then radiate a very broad band photon spectrum, extending from radio frequencies to multiTeV gamma-rays, through synchrotron and Inverse Compton processes.   

The accelerated pairs propagate far from the termination shock, advected together with the toroidal magnetic field lines, as an expanding MHD fluid. They are confined within the cavity escavated by the wind and progressively lose energy, as also demonstrated by the fact that the radio emitting region is typically broader than the X-ray emitting region, in agreement with a scenario in which propagation is energy independent and the difference in size is due to the larger energy losses suffered by the higher energy particles responsible for X-ray emission.

The question of whether some of these pairs can leave the system has acquired a special interest in recent times, after the discovery of a positron excess in the energy region 10-100 GeV by the PAMELA satellite \cite{Pampos}. The only guaranteed source of diffuse positrons in the Galaxy is represented by inelastic nuclear collisions of cosmic rays, which give rise to charged pions and in turn to positrons (and electrons) of secondary origin. The spectrum of such positrons is steeper than the cosmic ray spectrum, because of propagation and loss related effects. The PAMELA data show an increase of the $e^+/(e^++e^-)$ fraction with energy, that cannot be explained in the traditional scenario, and has been interpreted as the result of the intervention of an additional source of positrons. Several possibilities have been discussed in the literature: reacceleration of positrons originating as secondary products of CR interactions inside supernova remnants \cite{blasi,blasiserpico}, inhomogeneous distribution of sources in the solar neighborhood \cite{nir}, dark matter annihilation in the Galaxy \cite{cholis,sommer,grajek} and finally pulsars \cite{hooper,grasso}. Severe constraints on the origin of the PAMELA excess are imposed by the PAMELA observations of the antiproton \cite{PamPbar} flux and by the recent measurement of the electron spectrum by the Fermi satellite, up to $\sim 1$ TeV\cite{FermiEle}. The former suggest that whatever the source of the positrons is, it has to be less effective in producing antiprotons, at least up to $\sim 1$ TeV energy. The combination of Fermi and PAMELA results provide constraints on the spectrum of the positrons produced by the additional putative sources and on their spatial distribution. The case in which dark matter (DM) annihilation is assumed to be the main source of positrons offers an interesting demonstration of how powerful these constraints are: standard WIMPS would in fact result in a too low flux of positrons. In order to make DM annihilation a viable scenario, one is forced to assume either a boosting factor due to substructures and/or an enhancement of the annihilation cross section, as might be caused by the Sommerfeld effect \cite{sommer}. While the former is unlikely to provide an effect of more than a few, an enhancement of the annihilation rate also leads to a corresponding increase of the antiproton flux, in conflict with observations. If, again, one wants to force a DM interpretation, the DM candidate has to be assumed as leptophilic. Following reasonable scientific standards, this scenario can be considered highly disfavored if not ruled out. 

As stressed above, pulsars naturally produce electron-positron pairs. Therefore they have been immediately proposed as possible sources of the positrons observed by PAMELA \cite{hooper}. The main issue with pulsars as sources of cosmic ray positrons is in the possibility for the positrons to escape the supernova remnant environment. For this reason the authors of \cite{hooper} explicitly referred to mature pulsars as sources. The qualitative nature of the justification for considering mature pulsars opens however the way to criticisms for this scenario, strengthened by the fact that the attention was there centered on pulsars rather than pulsar winds, while the latter are the real place where the genesis of the final positron spectrum should be sought. 

Here we discuss the problem of the positron excess and of its possible connection with pulsars in the context of the much more complex phenomenology of pulsar winds and pulsar wind nebulae, on which we now have a wealth of data in different radiation bands. The paper is organized as follows: in \S~\ref{sec:inside} we briefly discuss the different parts of the relativistic wind launched by a pulsar inside a supernova remnant. In \S~\ref{sec:bow} we discuss the importance of bow shock nebulae as possible vehicles to carry positrons (and in general pairs) outside the remnant to make them available in the cosmic radiation. In \S~\ref{sec:excess} we describe the calculations of the positron flux at Earth as a result of injection from individual pulsars. We dedicate special attention to discussing the effect of discreteness in the Galactic pulsar distribution on the positron excess and on the spectrum of electrons (meaning here the sum of electrons and positrons) at Earth. Finally, \S~\ref{sec:discuss} is then devoted to a critical discussion of the achievements and challenges of the pulsar wind scenario for the positron excess. 

\section{A pulsar wind inside a supernova remnant}
\label{sec:inside}

In this section we discuss some selected topics which are of relevance for the problem of pair production in the pulsar environment and therefore for the problem of the positron excess in cosmic rays. For obvious reasons we will not provide here a full review of all the Physics involved in the production of pulsar winds and their interactions with the surrounding medium.  Excellent recent reviews of these issues can be found in \cite{arons09,gaeslane}. 

A rotating magnetized neutron star loses rotational energy. The decrease of the rotation frequency $\Omega=2\pi/P$ (where $P$ is the rotation period) can be written in general form as 
\be
\dot \Omega = -\alpha \Omega^{n},
\label{eq:omega}
\ee
where $n$ is the so-called braking index ($n=3$ for magnetic dipole). The spin down luminosity of the pulsar can easily be written in terms of the moment of inertia $I$:
\be
\dot E = I \Omega |\dot \Omega| = \alpha I \Omega^{n+1}.
\ee
Eq. \ref{eq:omega} can be integrated to give
\be
\Omega(t) = \frac{\Omega_{0}}{\left[1+\frac{t}{\tau_{0}}\right]^{\frac{1}{n-1}}}~~~~~~~\tau_{0}=\frac{1}{\alpha (n-1) \Omega_{0}^{n-1}},
\label{eq:tau0}
\ee
and the spin down luminosity becomes:
\be
\dot E = \alpha I \frac{\Omega_{0}^{n+1}}{\left[ 1 + \frac{t}{\tau_{0}}\right]^{\frac{n+1}{n-1}}}
\ee
The special case of magnetic dipole corresponds to $n=3$ and $\alpha=(5/8)B_{s}^{2}R_{s}^{4}/M_{s} c^{3}$, where $B_{s}$ is the surface magnetic field, $R_{s}$ is the radius and $M_{s}$ the mass of the neutron star. The total energy lost by the star is $(1/2) I \Omega_{0}^{2}\approx 10^{50}$ erg, most of which is converted into a magnetized, relativistic wind in a time comparable with $\tau_{0}$. 

The relativistic wind launched by the pulsar impacts on the much slower ejecta of the supernova that generated the pulsar in the first place. This creates a reverse shock that moves inward and a contact discontinuity is formed between the PWN material and the ejecta. This reverse shock is what is usually named {\it termination shock}. The nebulae that we observe are produced by non-thermal particles downstream of the termination shock where the incoming pairs are partly thermalized and, even more important, accelerated to very high energies (in the case of the Crab, radiation from particles with energies up to a few $\times 10^{15}$ eV is observed). 

The physical picture that describes the launching and the propagation of the pulsar wind from the star magnetosphere to the termination shock still presents a number of lose-ends. We will not touch these issues here, while extensive discussion can be found in \cite{arons09,kirk09}. What is relevant for our present goals is that: 1) the wind is mostly made of electron-positron pairs, while the presence of a numerically negligible, but energetically important fraction of ions cannot be excluded (see also below); 2) the wind must originate as a Poynting flux dominated outflow, but from observations of Pulsar Wind Nebulae we infer that at the termination shock most of its energy is in the form of particle kinetic energy; 3) at the termination shock the wind is highly relativistic, with a Lorentz factor $\Gamma$ in the range $10^4-10^7$, as derived again from observations and modeling of PWNe. The terminal Lorentz factor of the wind is crucially important in a number of respects for the constraints it puts on the physics of the pulsar magnetosphere. In the last few years, much has been learnt on the wind properties close to the termination shock thanks to extensive comparison between the results of 2-D relativistic MHD simulations of the pulsar outflow and observations of PWNe (\cite{kl04,ldz04,ldz06,volpi08}). Nonetheless, among all unknown parameters, the wind Lorentz factor is the most elusive to this kind of diagnostics, since the dynamics of the post-shock flow is independent of $\Gamma$ as long as $\Gamma\gg 1$. At present, the only viable means to put constraints on it resides in the modeling of the global emission spectrum of Pulsar Wind Nebulae: investigations have only just started \cite{nic10}. 

The evolution of the Pulsar Wind Nebula depends on the properties of the surrounding supernova ejecta. In the simplest picture one can assume that the supernova ejecta form a spherical shell around the pulsar. In this case, during the ejecta dominated stage the wind expands in a quasi-spherical way and drives a shock in the ejecta. When the supernova enters its Sedov phase, the reverse shock driven by the blast wave reaches the inner part of the remnant and in principle excites a few reverberations in the nebular structure \cite{van01,nic03}. On the other hand, by the beginnig of the Sedov phase the kick birth velocity of the pulsar, whose distribution shows a peak at $\sim 400-500$ km/s \cite{kicks02}, is expected to have forced the pulsar off center, so that the structure of the nebulae in this phase is likely rather asymmetrical. The issue of the effect of the reverse shock on the population of electron-positron pairs in terms of adiabatic compression and/or spatial displacement, would be worth being investigated in more detail. 

In order to establish, if possible, a connection between Pulsar Winds and the positron flux in Cosmic Rays, what is really important for us is the pair spectrum that is released in the ISM and the only way to obtain some insights on this issue is through observations of PWNe. As stressed above, all radiation observed in PWNe comes from particles radiating downstream of the termination shock, where the pairs initially produced in the pulsar magnetosphere are accelerated to non-thermal energies. On purely observational grounds, the spectrum of radiation observed from several PWNe requires a spectrum of radiating electrons and positrons which has the shape of a broken power law with a break at Lorentz factor of $\sim 10^{5}$, namely $E_{b}\sim 50$ GeV for electrons \cite{nic10}. The slope below this point is typically $\alpha_{1}\approx 1-1.5$, while the slope at high energies is $\alpha_{2}\approx 2-2.4$. This seems to be a rather general trend, and as we discuss in the next section, appears to be common also to fast pulsars outside their SNRs (bow shock nebulae). 

The acceleration process responsible for the production of this spectrum is all but clear, for many good reasons. First the broken power law is not typical of stochastic acceleration processes, in which usually a single power law spectrum is produced (or in some cases a curved spectrum, rather than two power laws). Second, any stochastic acceleration process that we are aware of leads to the production of a non-thermal tail on top of a thermal distribution of particles which take into account the unavoidable heating process. This is certainly true in the case of shock acceleration: at the relativistic termination shock one would expect to have a thermal distribution corresponding to particles with mean energy $\sim (\Gamma-1) m_{e}c^{2}$, where $\Gamma$ is the Lorentz factor of the wind, and a non-thermal tail in the high energy part. No evidence has ever been found of this thermal component in PWNe, which makes the situation really puzzling. One could speculate that the thermal ``bump'' is at lower energies, where either it is not observed or cannot be observed (for instance if its energy corresponds to frequencies of the synchrotron photons below the ionospheric cutoff of the atmosphere). This possibility however opens more theoretical problems than it solves: for an electron spectrum $N(E)\propto E^{-\alpha_{1}}$, with $\alpha_{1}<2$ as observed, most particles are in the lowest energy end of the distribution, while most energy is at the high energy end. Moving the thermal peak at very low energies implies that the number of pairs reaching the termination shock is much larger than can be accounted for by current theoretical models of pair cascading in the pulsar magnetosphere. 

While the slope of the spectrum at $E>E_{b}$, $\alpha_{2}\sim 2-2.4$, appears to be similar to that expected from diffusive shock acceleration at a relativistic shock, a closer look reveals more problems: the termination shock is a highly relativistic quasi-perpendicular shock, and in these circumstances simple physical arguments lead to think that shock acceleration cannot take place efficiently, as also confirmed by PIC simulations \cite{sironi09}, which have never shown evidence of acceleration in perpendicular shocks occurring in pure pair plasmas if the magnetization is at the level implied for most of the pulsar outflows \cite{volpi08}. Moreover, diffusive shock acceleration could not explain, even in principle, very flat low energy spectra $\alpha_{1}<2$, with the possible exception of the so-called large angle scattering regime \cite{vietri,baring}, but there are serious doubts that the conditions for large angle scattering may be actually realized in Nature.

In principle, acceleration by cyclotron absorption \cite{hoshino92,amato06} could produce the desired spectra if a substantial fraction of the pulsar wind energy is carried by protons. However, this condition can only be realized if the wind Lorentz factor is sufficiently high: it is indeed difficult to imagine that the pulsar wind can carry a proton density in excess of the Goldreich-Julian value, so in order for them to be energetically dominant the wind Lorentz factor needs to be sufficiently high and the pair multiplicity relatively low ($\Gamma\sim 10^6$ and $\kappa\sim 10^3$ in the case of Crab). The requested values of $\Gamma$ and $\kappa$ agree with current theoretical modeling of pair production but are at odds with the inferences of the most recent studies on the subject \cite{nic10}. In any case this scenario would leave the issue of the thermal peak untouched. 

Another scenario that can be considered to explain the observed spectra is that arising from a picture in which the striped magnetic structure \cite{coroniti90,lyubkirk01} implied for the wind at low latitude around the pulsar rotational equator leads to reconnection at the termination shock, with associated particle acceleration \cite{lyub08}. The shape of the spectrum is poorly constrained in this case, due to its strong dependence on the geometry of the reconnection region. A priori it is however difficult to envision a reason to have a broken power law. On the other hand, the arguments to deduce that there should be a thermal component downstream of the shock might be weaker in this case, since energy could be transferred to bulk motion of the plasma rather than disordered (thermal) energy (Jon Arons, private communication).

\section{A pulsar wind escaping the parent supernova remnant: bow shock nebulae}
\label{sec:bow}

The typical birth velocity of a pulsar born in a core collapse supernova event is $\sim 400-500$ km/s \cite{kicks02}, while about 50\% of the pulsars have velocities larger than this. This fact might be of the highest relevance for the problem of the escape of electron-positron pairs from the pulsar environment, since a fast pulsar can release a large part of its energy through a relativistic pair wind that propagates directly in the ISM, where the confining effect of the ejecta is absent. After the pulsar is born, its proper motion takes it far from its birth place, across the central part of the remnant and later across the shocked hot ejecta. If we assume that the SNR is in its Sedov phase, the radius of the blast wave is traced by the relation:
\be
R_{sh}(t)=R_{S} \left( \frac{t}{t_{s}}\right)^{2/5},
\ee
while the pulsar covers a distance $R_{pulsar}=v_{kick}t$. It is straightforward to see that, for the typical values of $v_{kick}$ quoted above, the pulsar leaves the parent supernova remnant after a time of order $\sim 40,000-50,000$ years (the assumption of Sedov expansion is justified for reasonable values of the supernova and ISM parameters). The motion of the pulsar in the ISM is at this point supersonic, therefore the impact of the relativistic wind on the ISM leads to the formation of a bow shock structure. The forward (bow) shock also drives a reverse shock (termination shock), as illustrated in a schematic way in Fig.~6a of Ref. \cite{gaeslane}. At the termination shock particle acceleration leads to the formation of a non-thermal spectrum of pairs which produce a nebula, now referred to as {\it bow shock nebula}. The pairs which are accelerated at the termination shock (though, as discussed above, the acceleration mechanism is not known) propagate in the tail \cite{nic05} and eventually end up in the ISM in the form of cosmic rays. The spectral distributions of accelerated pairs required to explain the spectra of observed radiation, in the few cases in which observations allow such a study, are very similar to those observed in PWNe inside SNRs. For instance in G319.9-0.7, the bow shock nebula powered by PSR J1509-5850, the radio spectrum \cite{psrj1509} has a slope $-0.26$ (corresponding to an electron spectrum with slope $\alpha_{1}=1.52$); in the {\it Mouse} \cite{mouser} the radio spectrum has slope $-0.3$, corresponding to $\alpha_{1}=1.6$. The high energy part of the spectrum is also similar to that observed in PWNe inside SNRs \cite{mousex}.

The fact that pulsars can exit their birth site after 40-50 kyr is clearly of paramount importance in terms of establishing a possible connection with the cosmic ray positron flux detected by PAMELA. We investigate this possibility by first estimating the rotational energy left in the pulsar at the time of escape from the remnant. We carry out this estimate keeping the braking index arbitrary to start with. The rotational energy left after a time $T_{e}$, which we take as the escape time from the remnant, can be written as
\be
E_{rot}(t>T_{e}) = \frac{1}{2} I \Omega_{0}^{2} \left( 1 + \frac{T_{e}}{\tau_{0}}  \right)^{\frac{2}{1-n}}.
\label{eq:avail}
\ee
It is easy to see that the critical parameters to estimate how much energy the pulsar can still convert into electron-positron pairs after its escape from the remnant are the initial rotation frequency $\Omega_{0}$ and the characteristic time $\tau_{0}$ (which in turn depends on $\Omega_{0}$ and on the braking index $n$ as in Eq. \ref{eq:tau0}). These parameters are directly measured only in the very rare cases in which one has observations spanning a long enough period of time so as to allow derivation of all three quantities $\Omega$, $\dot \Omega$ and $\ddot \Omega$. Using the expressions reported in the previous section one can then easily obtain the braking index as 
\be
n=\frac{\Omega \ddot \Omega}{\dot \Omega^{2}}.
\label{eq:break}
\ee
Moreover the characteristic time $\tau_{0}$ is such that
\be
\tau_{0}+T_{age} = \frac{1}{1-n}\frac{\Omega}{\dot \Omega},
\ee
where $T_{age}$ is the age of the pulsar at the time of measuring the values of $\Omega$, $\dot \Omega$ and $\ddot \Omega$. The difficulty in measuring all these rotational parameters justifies the fact that the full set of spin down parameters is known so far for only four pulsars, the Crab pulsar \cite{crabn}, Vela \cite{velan}, B1509-58 \cite{B1509n} and B0540-69 \cite{B0540n}. In all these cases the measured braking index $n$ is lower than in the dipole case ($n=3$), ranging from the very low value of Vela ($n=1.4$) to about 2.8 for B1509-58. 

For the sake of simplicity we consider two benchmark cases, a hypothetical dipole case ($n=3$) with initial rotational period $P_{0}=10$ ms (corresponding to $\Omega_{0}=628.3~\rm s^{-1}$) and a Crab-like case, with $n=2.5$ and $P_0=19$ ms, as deduced from observations. 

For the dipole case, one easily obtains that $\tau_{0}\approx 5000$ years. In this case, if the escape time is estimated as $T_{e}\simeq 4\times 10^{4}$ years, from Eq. \ref{eq:avail} one gets the energy available for conversion into pairs after the pulsar escapes the remnant: $E_{rot}(t>T_{e}) \approx 2\times 10^{49}$ erg. If one calculates the time $\tau_{0}$ from observations, the result is quite different. The observed values of the spin down parameters are as follows: $\Omega\approx 190 s^{-1}$, $\dot \Omega=-3.86\times 10^{-10}s^{-2}$ and $\ddot \Omega = 1.24\times 10^{-20}s^{-3}$. Using Eq. \ref{eq:break} one finds $n=2.51$ and $\tau_{0}\approx 730$ years (a factor $\sim 7$ shorter than for the dipole case) taking into account the age of the Crab pulsar (or using the measured value of the third time derivative of $\Omega$, as in \cite{crabn}). This approach also returns $\Omega_{0}\approx 330 s^{-1}$. In this case, the energy available after escape is $E_{rot}(t>T_{e}) \approx 3\times 10^{47}$ erg. The different conclusions in the two cases show how the results depend in a crucial way on the spin down history of the pulsar: the efficiency in conversion to pairs needed to explain the PAMELA data are also very different as we discuss in the section below. The dipole case would be even more at odds with the observed data if one assumed the same value of $\Omega_{0}\approx 330 s^{-1}$ in the two cases.

The positrons produced by a pulsar after the escape from the SNR can hardly be confined by the bow shock. The nebulae that we observe have a morphology that suggests that relativistic particles are being advected far downstream of the pulsar, while the bow shock opens up. It is therefore only reasonable to envision that these positrons can escape the nebula and be part of the cosmic radiation propagating in the Galaxy and eventually reach the Earth. This addresses one of the main concerns that have been raised about a pulsar related origin of the PAMELA positrons, namely that of the leakage of such positrons outside the parent remnant. 

On the other hand, the issue of the fate of the positrons produced while the pulsar is inside the parent remnant remains there. A dedicated investigation of this problem, even independently of the positron signal, would be worthwhile.

\section{The positron flux from pulsars}
\label{sec:excess} 
 
The flux of positrons (and electrons) from pulsars has been calculated and discussed in previous papers \cite{pulsar} and compared with the flux of positrons observed by PAMELA \cite{hooper,grasso}. These calculations have however two limitations: 1) they all start with a {\it ad hoc} assumption on the spectrum of positrons at the sources, without a proper discussion of observational bounds and physical interpretation of the acceleration and escape of these particles. 2) The pulsars are assumed to represent a population of sources distributed continuously (in space and time) in the Galaxy, namely the discrete nature of the sources is ignored (though in \cite{grasso} the authors discuss the role of known nearby pulsars). 

In this section we carry out the calculation of the positron and electron fluxes at the Earth assuming that they are accelerated in individual pulsar winds and propagate from each source to the Earth through the Galactic magnetic field. At the same time, we assume that ordinary cosmic rays (protons and electrons) are accelerated in supernova remnants associated with the formation of pulsars. This approach neglects the contribution of supernovae of type Ia, which are subdominant in our Galaxy. Their contribution will be discussed elsewhere. The calculation takes into account the temporal evolution of the distribution function of electrons and positrons. This allows us to offset the production time of positrons from a pulsar by the time required for the pulsar to escape the remnant after the time of explosion of the parent supernova. Moreover, the energetics available in the pulsar wind at that time is calculated as discussed in the previous section, taking into account the spin down of the pulsar before the escape from the remnant. 

The time and location of each supernova event in the Galaxy are drawn at random from the distribution of type II supernovae as given in \cite{ferriere}. Once time and location are known with respect to the location of the Sun, the diffusion equation is solved by using the appropriate Green function. In the range of energies we are interested in, the spectrum of the produced electrons and positrons is affected by propagation mainly through radiative losses, therefore for simplicity we neglect here the role of escape of electrons from the Galaxy, which play a role only below 10 GeV. In this case, the solution of the diffusion equation for an individual supernova remnant is \cite{syro59}: 
\be
n(E,t) = \int_{E}^{\infty} dE' \frac{\exp\left[ -(\vec r - \vec r_{exp})^{2}/4\lambda(E,E') \right]}{|b(E)| (4\pi \lambda(E,E'))^{3/2}} N(E') \delta(t-t_{exp}-\tau(E,E')),
\ee
where 
\be
\lambda(E,E') =  \int_{E'}^{E} dy \frac{D(y)}{b(y)}~~~~~\tau(E,E') = \int_{E'}^{E} dy \frac{1}{b(y)}.
\ee
Here $D(E)$ is the diffusion coefficient experienced by electrons and positrons while propagating throughout the Galaxy and $b(E)=-AE^{2}$ is the rate of energy losses due to synchrotron and inverse Compton scattering (ICS). The time $t_{exp}$ is the time of explosion of the supernova in the case of electrons, while it is the time of escape of the pulsar in the case of production of electrons and positrons in the relativistic wind of the pulsar. The distribution function $N(E)$ is the injection spectrum of either the SNR or the pulsar wind. In both cases the injection is assumed to be instantaneous. In the case of pulsars, the assumption is well justified because the spin down of the pulsar concentrates most injection at early times, close to the time of escape of the pulsar from the remnant. For SNRs, which in our picture only produce electrons, the injection of accelerated particles is not well understood (see for instance \cite{escape} for a detailed discussion of the issues involved in the CR escape from a SNR). However the effects of the assumption of instantaneous injection on the results have been tested and found to be minimal at all energies of interest for the positron production.

The contribution of all pulsars is calculated by exploding Supernovae and taking into account the associated SNRs and pulsars for a period exceeding all relevant time scales in the problem (for instance escape time from the Galaxy and time of energy losses), so as to ensure that the galaxy is globally in a stationary situation. This does not mean that the flux of electrons and positrons at the Earth is stationary. In fact, proximity effects lead to a time-dependence of the fluxes and, even more important, to a dependence of the resulting fluxes upon the particular realization of the distribution of supernovae in the Solar neighborhood. The diffusion coefficient is assumed to be $D(E)\propto E^{1/3}$ to avoid problems with CR anisotropy in the energy region $10^{4}-10^{6}$ GeV. The normalization of $D(E)$ is taken so as to satisfy the condition that the escape time of CRs at energy of $10$ GeV/nucleon is $\sim 1.5\times 10^{7}$ years, as deduced from the measurement of the B/C ratio \cite{BC}. This normalization depends of course upon the size of the halo $H$ from which particles have to escape diffusively (all our calculations are carried out for $H=2$ kpc).

From the point of view of escape of a pulsar from the SNR, we consider two benchmark cases: one in which the pulsar spins down as a magnetic dipole, and the other one in which the spin down occurs with a braking index $n=2.5$. The latter case is closer to observations, if we assume that the four pulsars for which the rotation parameters could be measured are 'typical'. The implications of these two models for the pulsar wind scenario of origin of the positron excess are, as we show below, rather impressive. 

It is important to realize that the approach presented here returns automatically the spectra of both primary electrons accelerated in SNRs and electrons and positrons produced in pulsar winds after the escape of the pulsar from the SNR. Actually the same approach (completed with the escape of cosmic rays from the Galaxy) also returns the spectrum of CR nuclei at the Earth, as discussed in \cite{abZ}. The authors also reach there the conclusion that the anisotropy of CRs at the Earth exceeds the observations if the diffusion coefficient has an energy dependence around $\sim E^{0.6}$.

Before illustrating the results of detailed calculations, it is worth discussing briefly some general points concerning the contribution of individual SNRs and pulsar winds to the spectrum of electrons and positrons at the Earth. For the sole purpose of this simple estimate, we assume for simplicity that the sources are distributed uniformly in the disc of the Galaxy, with a half-thickness $H_{d}=150$ pc and a radius $R_{g}=15$ kpc. The sources that can contribute to the flux at Earth at a given energy $E$ are those lying within a distance from the Earth such that the propagation time is shorter than the loss time at that energy. 
If $\cal R$ is the rate of supernovae in the Galaxy (assumed to be one every 30 years), the number of supernovae within a given distance R from the Earth, exploding in a loss time, is ${\cal R} \tau_{loss} (R/R_{g})^{2}$. For our purposes: $R\approx (4 D(E) \tau_{loss})^{1/2}$. It follows that the number of sources which can contribute to the flux at energy $E$ is 
\be
{\cal N} \approx {\cal R} \frac{4D(E)}{R_{g}^{2}} \tau_{loss}^{2} \propto E^{1/3}/E^{2}.
\ee
This simple estimate shows that the number of relevant sources is a rapidly decreasing function of energy. For typical values of the parameters $\sim 1000$ sources contribute at $10$ GeV, and $\sim 20$ at $100$ GeV. At $1$ TeV only about 1 source can contribute. This conclusion has several implications: 1) the flux of electrons and positrons at high energy is a sensitive function of the local distribution of recent SNRs; 2) the slope of the electron spectrum as inferred from the standard stationary solution of the transport equation is typically not reproduced in a time dependent approach with discrete point sources exploding at random times. 

\begin{figure}[t]

\includegraphics[scale=.35]{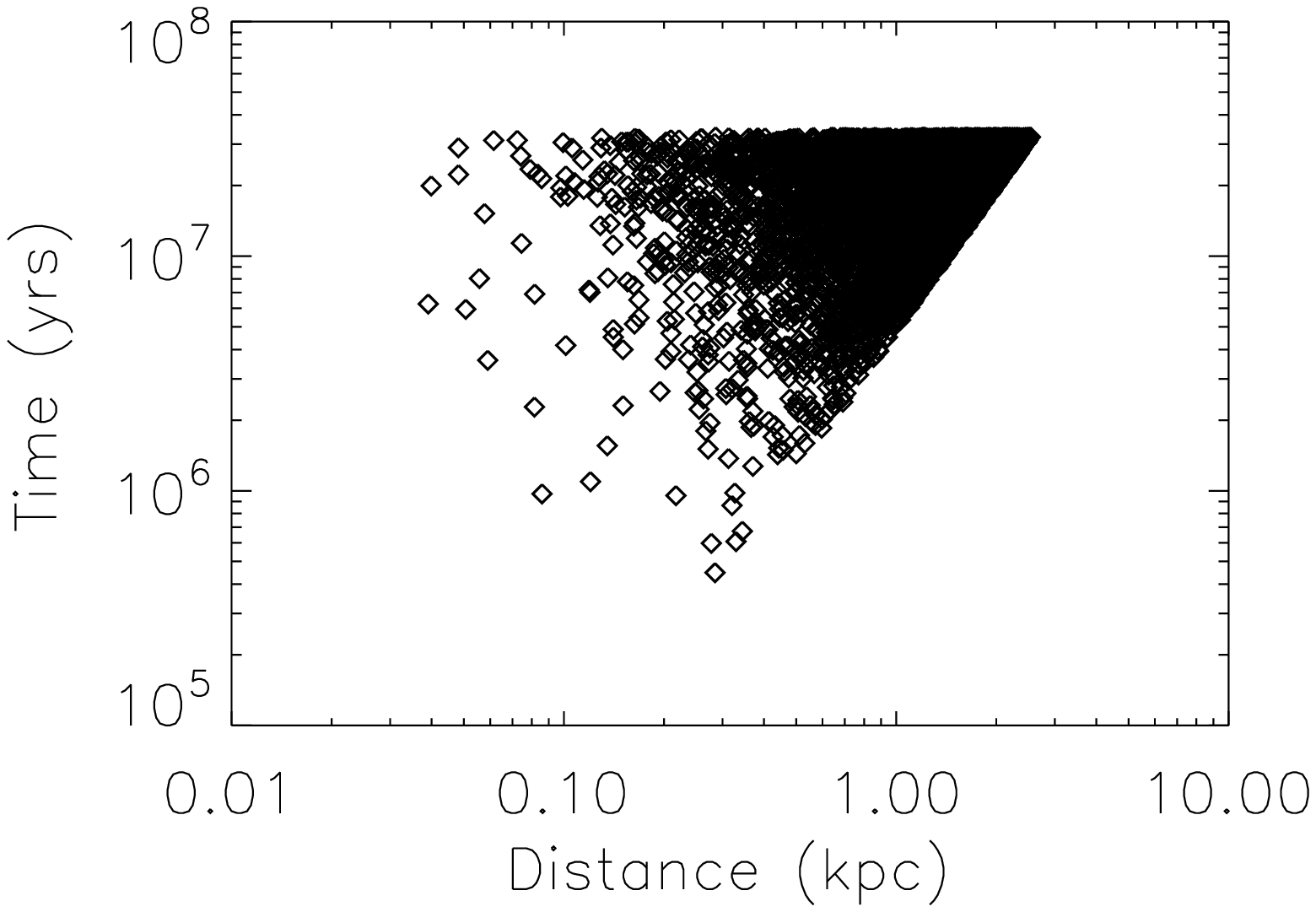}
\includegraphics[scale=.35]{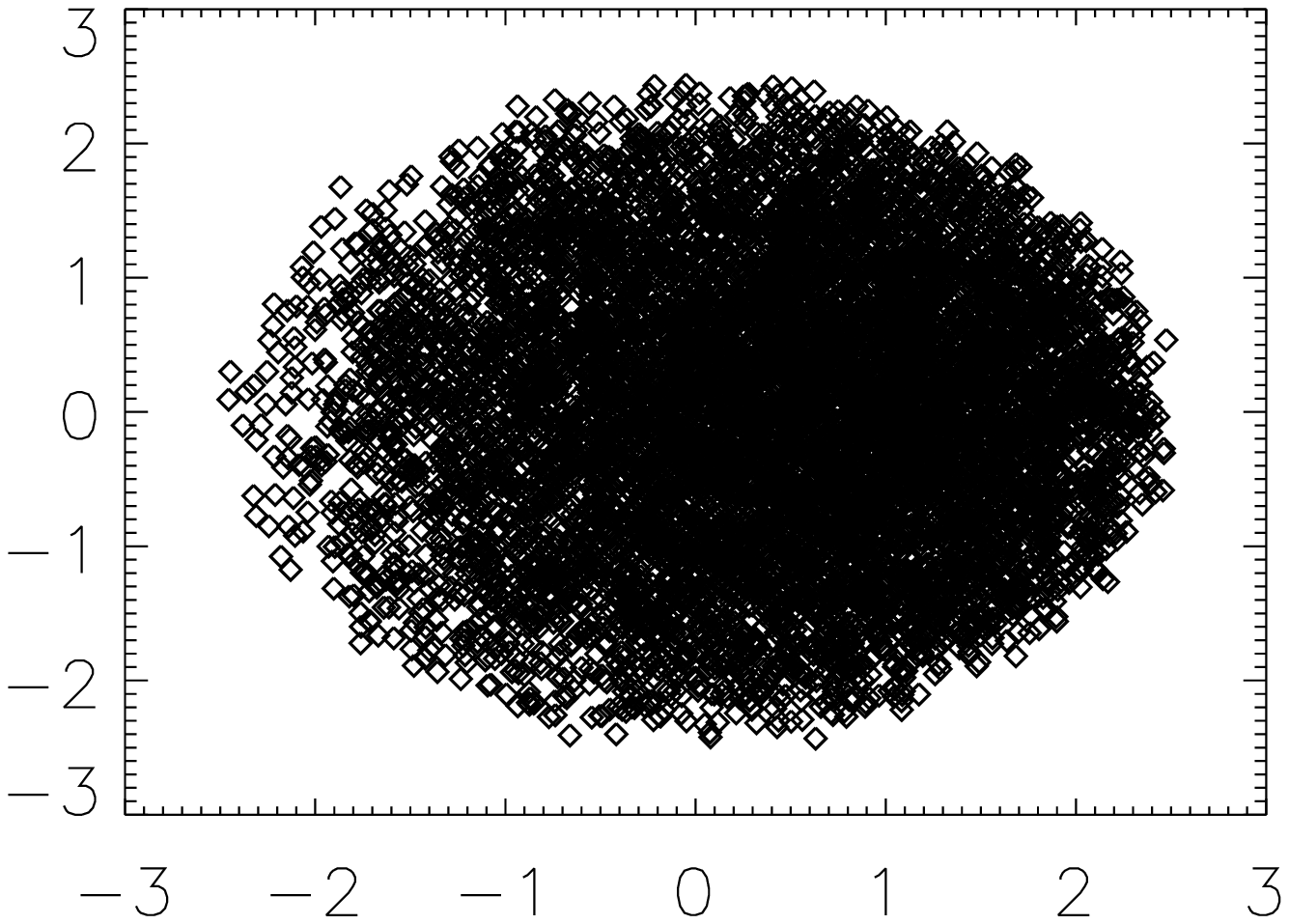}

\includegraphics[scale=.35]{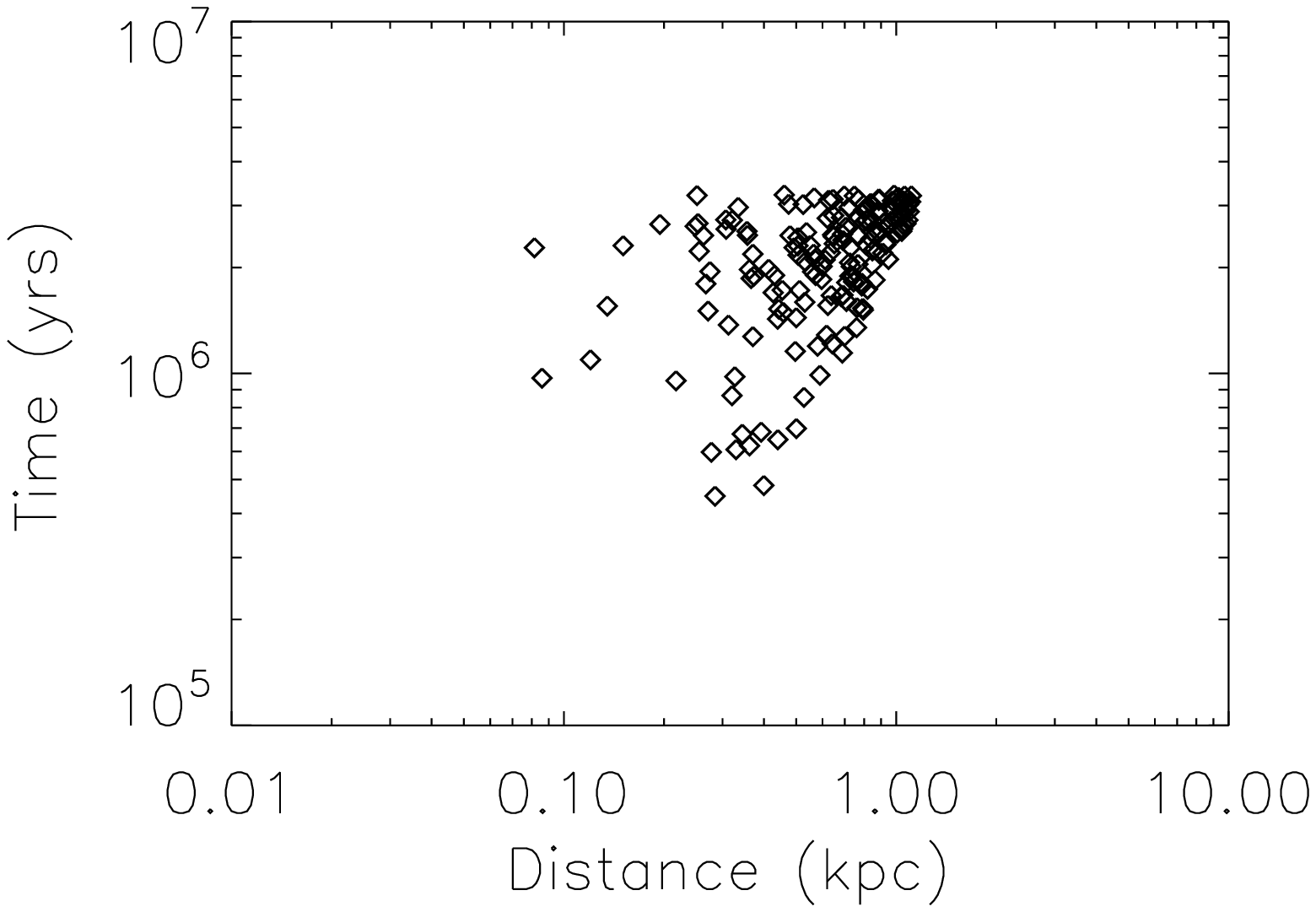}
\includegraphics[scale=.35]{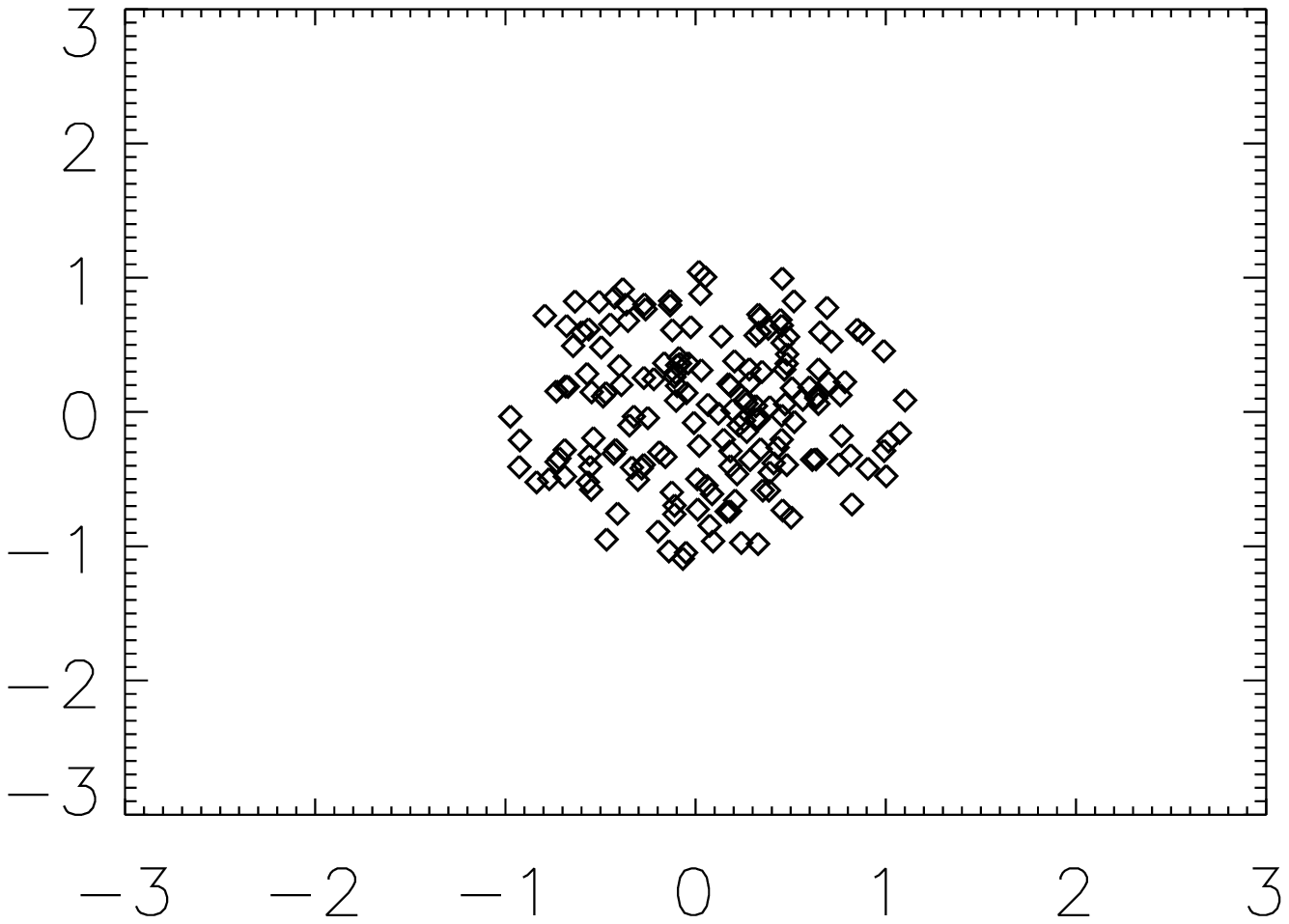}

\includegraphics[scale=.35]{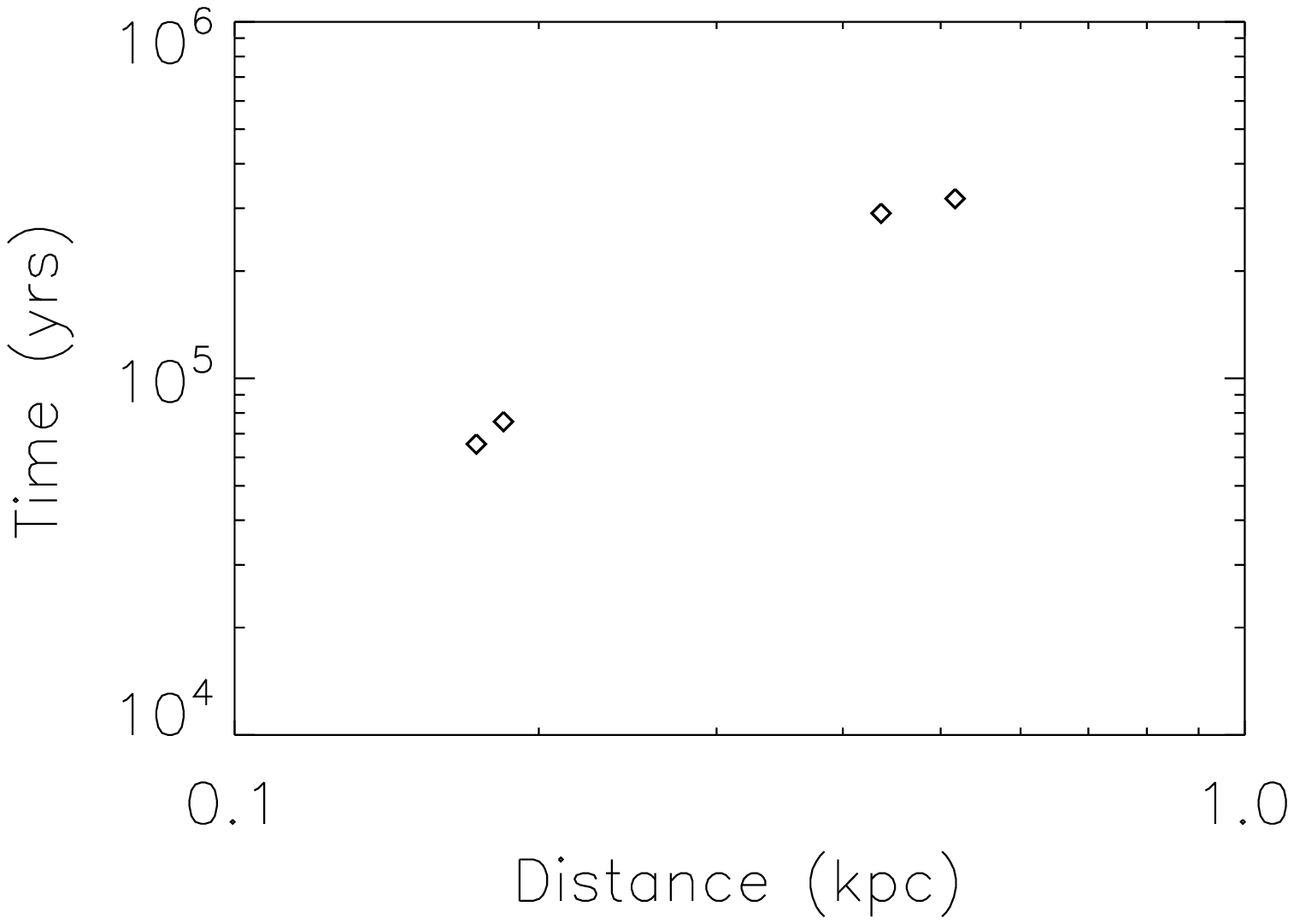}
\includegraphics[scale=.35]{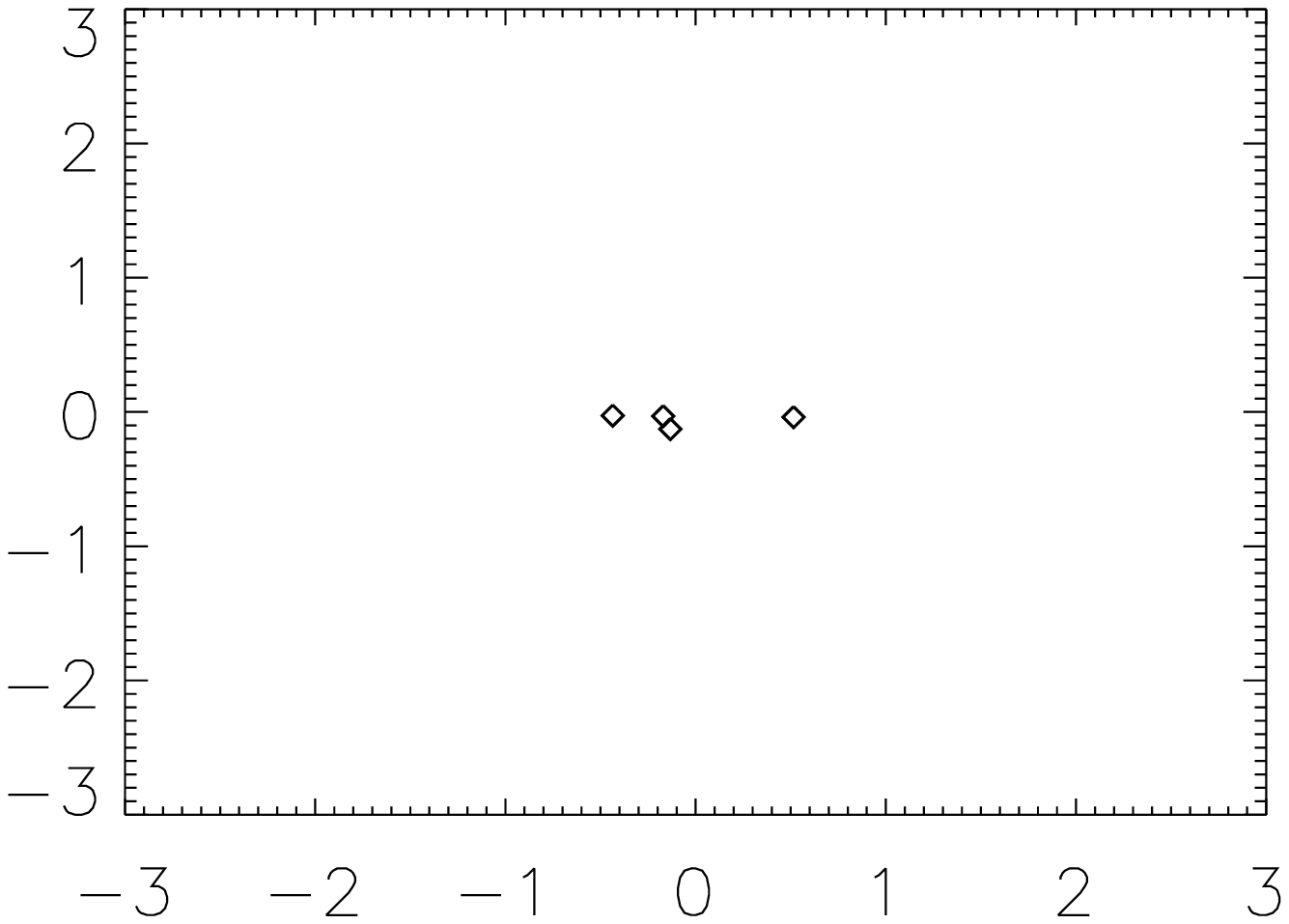}

\caption{Temporal and spatial distribution of sources in one realization as a function of energy. {\it Left Panels)}: Time of explosion of a supernova as a function of the distance from the Sun. {\it Right Panels)}: Cartesian spatial coordinates (in kpc) of the supernova events on a grid centered on the location of the Sun. The first row refers to electron energy $10$ GeV, the second to $100$ GeV and the third to $1$ TeV.}

\label{fig:Nsource}       
\end{figure}

The distribution of type II supernovae \cite{ferriere} is not uniform across the disc of the Galaxy and shows a gradient along the radial direction. Using this distribution, it is possible to determine the number of SNRs/pulsars contributing at given energy in a more realistic way. The results are shown in Fig.~\ref{fig:Nsource} for energies of $10$ GeV (first row), $100$ GeV (second row) and $1$ TeV (third row). The plots on the left show the time of explosion (on the y-axis) as a function of the distance from the Sun location, while the plots on the right show the location of the sources in a given realization (the distance scale is centered on the Sun location). One can clearly see how increasing the energy the number of relevant sources (in space and time) decreases rapidly. In the 10 GeV case one may notice the asymmetric distribution of local supernovae as follows from the off-center location of the Sun in the Galaxy. 

It is worth stressing that the sources in these plots do not contribute all in the same way: there are nearby sources which turned on a long time ago, while there are more distant sources which are recent and may contribute more. In other words, the flux at Earth is in general a complex combination of the contributions from these sources. Only for a truly continuous (in space and time) distribution of the sources one recovers the standard solution of the stationary transport equation. 

We discuss now the results of the calculations of the flux of electrons (from acceleration in SNRs) and electrons and positrons coming from pulsar winds after the pulsar escapes the parent remnant. As mentioned above, we consider two models for the spin down of the pulsar, one in which the pulsar rotation slows down as a magnetic dipole (braking index $n=3$) and one in which the spin down corresponds to a braking index $n=2.5$, similar to the one observed for the Crab pulsar. In both cases the average escape time from the remnant has been taken to equal $4\times 10^{4}$ years. 

\begin{figure}[t]

\includegraphics[scale=.65]{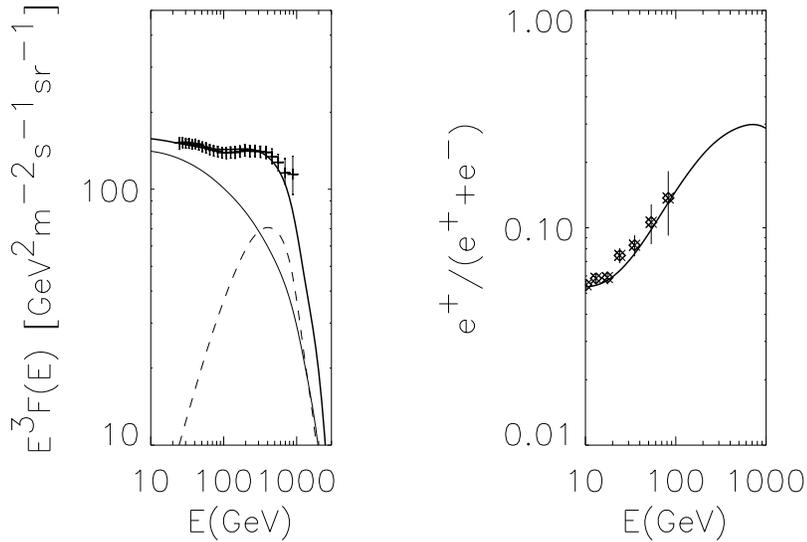}

\caption{Case of dipole spin down with escape of the pulsar from the remnant 40,000 years after the supernova event. {\it Left)}: Spectrum of electrons from SNRs (solid line) and electrons and positrons from pulsar winds (dashed line). The thick solid line through the Fermi data points is the sum of the two. {\it Right)}: Positron ratio compared with the PAMELA data points.}
\label{fig:dipole}  
\end{figure}

When the spin down is dipole-like, the predicted spectrum of electrons and positrons and the positron ratio are illustrated in Fig.~\ref{fig:dipole}. The data points in the left panel are from the Fermi satellite \cite{FermiEle}, while the data points on the right panel are from the PAMELA satellite \cite{Pampos}. The acceleration efficiency of protons is $\sim 10\%$, and the ratio of the accelerated spectra of electrons and protons is $K_{ep}=9\times 10^{-3}$. The injection spectrum of electrons, which are still assumed to be mainly accelerated via diffusive shock acceleration, together with protons, in Supernova blast waves is $\propto E^{-2.4}$ with a cutoff at $10$ TeV. The pairs from pulsar winds have an injection spectrum $\propto E^{-1.2}$ with a cutoff at $700$ GeV, as implied by observations of synchrotron radiation from PWNe in the relevant energy range. It is worth recalling that, as found in the previous section, the energy available in the pulsar rotation in the dipole case, after escaping the remnant, is of order $\sim 2\times 10^{49}$ erg. In this case the efficiency of escape of pairs from the nebula required to explain the positron excess and at the same time provide a good fit to the observed electron (plus positron) spectrum is $\xi_{\pm}=9\times 10^{-3}$, corresponding to $\sim 1.8 \times 10^{47}$ erg per pulsar wind. This estimate is similar to previous estimates appeared in the literature. 

It is worth stressing a few interesting things: 1) These predictions, though qualitatively similar, are quantitatively different if one changes the realization of sources, for the reasons explained above. We will discuss below the dependence upon realization. 2) If the discrete nature of the source is ignored, one would expect that electrons injected with a power law spectrum $E^{-\gamma}$ in a thin disc and allowed to propagate with diffusion coefficient $D(E)\propto E^{\delta}$ have an equilibrium spectrum $n(E)\propto E^{-\gamma}\tau_{loss}(E)/(4 D(E) \tau_{loss}(E))^{1/2} \sim E^{-3.06}$ for $\gamma=2.4$ and $\delta=1/3$. One can clearly see from the left panel of Fig.~\ref{fig:dipole} that this is not the case, as a consequence of the local distribution of sources around the Sun. 

A good fit to the electron spectrum measured by Fermi is possible only because of the superposition of the contributions of SNRs and pulsar winds. This superposition naturally explains the presence of wiggles in the electron spectrum and is the very reason for the PAMELA positron excess. 

\begin{figure}[t]

\includegraphics[scale=.65]{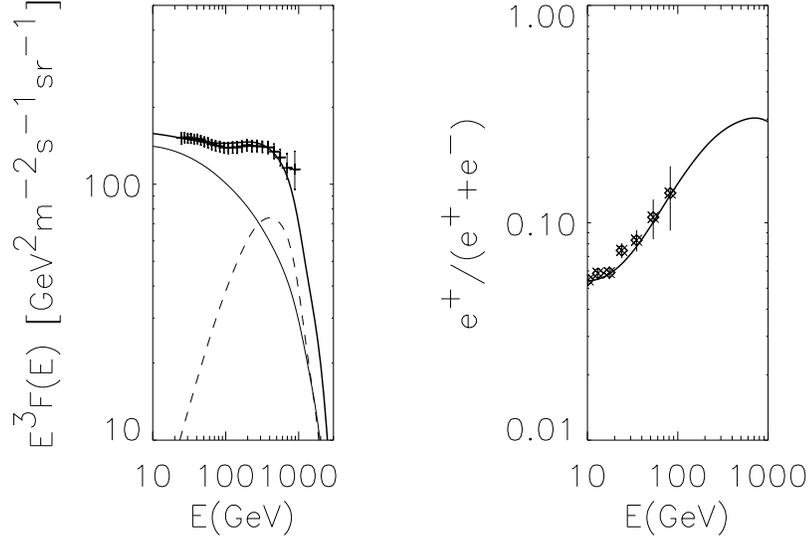}

\caption{Case of spin down with braking index $n=2.5$, with escape of the pulsar 40,000 years after the supernova event. {\it Left)}: Spectrum of electrons from SNRs (solid line) and electrons and positrons from pulsar winds (dashed line). The thick solid line through the Fermi data points is the sum of the two. {\it Right)}: Positron ratio compared with the PAMELA data points.}
\label{fig:n25}     
\end{figure}

An excellent fit to the data can also be obtained if one assumes a pulsar spin down with braking index $n=2.5$, as shown in Fig.~\ref{fig:n25}. In this case the parameters have the same values as in the dipole case, but the required efficiency for pairs in pulsar winds is $\sim 28\%$ (to be compared with the previous value of $\sim 9\times 10^{-3}$). It is worth stressing that this efficiency does not refer to the total rotational energy of the pulsar, but only to the energy available after the escape of the pulsar from the remnant. In this sense it does not sound as problematic that the required efficiency is of order tens of percent, since after the pulsar escapes the remnant high energy electrons can be hardly confined inside the nebula and they can only become part of the cosmic radiation, thereby contributing to the positron flux at Earth. 

\begin{figure}[t]

\includegraphics[scale=.65]{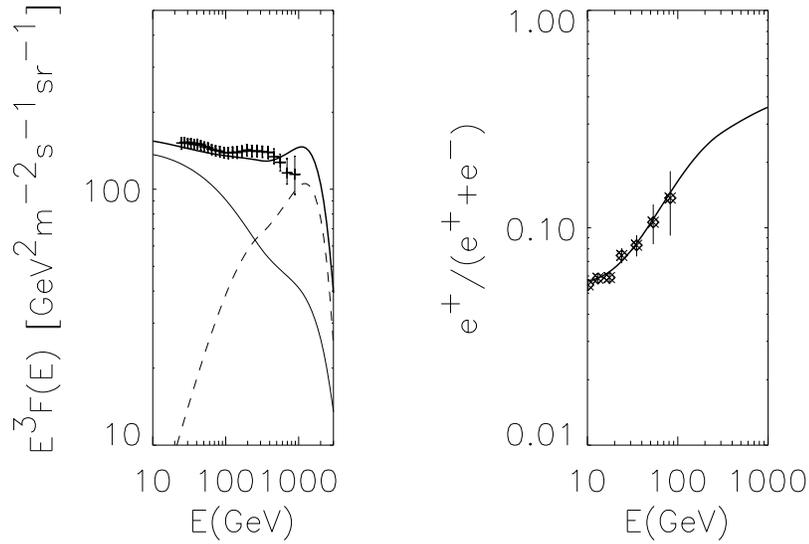}

\caption{Same as in Fig.~\ref{fig:n25} but for a different realization of source distribution in space and time.}
\label{fig:real}
\end{figure}

As anticipated above, the fact that sources are discrete in space and time leads to an intrinsic dependence of the predictions on the realization. In order to illustrate this dependence, in Fig.~\ref{fig:real} we show the results obtained for $n=2.5$ in a different realization of sources. One can see that in both contributions, of SNRs and pulsars, there is an excess at high energies, suggesting that in this realization a supernova happened to explode close to us in recent times. One can also see that the total flux is affected in the same way, but the positron ratio, which is measured at lower energies, remains basically unchanged, though future measurements with AMS2 might detect this type of proximity effects even in the positron fraction. 

\begin{figure}[t]

\includegraphics[scale=.65]{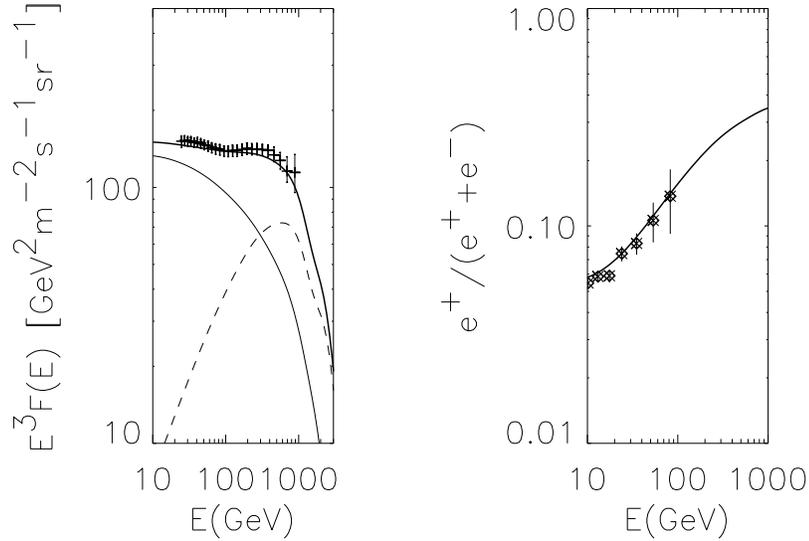}

\caption{Same as in Fig.~\ref{fig:n25} but with slope of the injection spectrum of pairs $-1.5$.}
\label{fig:sl15}      

\end{figure}

As we discussed in \S~\ref{sec:inside} and \S~\ref{sec:bow}, observations show that in the few cases in which we can measure the spectrum of the radiation and convert it to an electron spectrum, the low energy part has slope between $-1$ and $-1.8$. In order to illustrate the dependence of the predictions on the assumed injection spectrum of pairs in pulsar winds, we repeat our calculations for an injection spectrum $\sim E^{-1.5}$. The results are illustrated in Fig.~\ref{fig:sl15}, where the spectrum of pairs is cut off at $1.8$ TeV. The fit to Fermi and PAMELA data requires $K_{ep}=8.5\times 10^{-3}$ and an efficiency of conversion of rotational energy into accelerated pairs of $\xi_{\pm}=39\%$. Again, though quantitatively slightly different, qualitatively the conclusions are not excessively affected by the slope of the spectrum of electrons in the range of relevance for us. 

\section{Discussion}
\label{sec:discuss}

The discovery by the PAMELA satellite of a positron flux in excess of the secondary positron flux produced by inelastic CR interactions in the Galaxy has attracted much attention for many reasons. First, it represents one of the few macroscopic anomalies in our understanding of the origin of galactic cosmic rays; second, it has been viewed by some as one possible evidence of annihilation of dark matter; third, it inspired some to search for astrophysical sources of positrons; fourth, it happened at roughly the same time at which the Fermi satellite provided an accurate measurement of the electron spectrum (meaning the spectrum of electrons plus positrons), that showed interesting wiggles and a shape not immediately reconcilable with the predictions of the standard model of CR propagation in the Galaxy (though one should admit that such standard model is not that standard after all). 

It has also revived the interest of the scientific community for pulsars and pulsar winds as galactic antimatter factories. The connection between pulsars and the PAMELA positron excess has been almost immediate \cite{hooper}, but not much attention was devoted to the actual physics of the electron-positron pairs acceleration and escape from the environment where the pulsar is produced. 

In this paper we discussed the strong and weak points of the theory of pair generation in pulsar winds and their mixing with the cosmic radiation eventually observed at Earth after propagation in the Galaxy. We argued that the PAMELA positron excess is well explained if the pairs are liberated after the pulsar, originated in a core collapse supernova, escapes the parent supernova remnant. This event typically occurs $(4-5)\times 10^{4}$ years after the initial blast, as could be estimated by assuming an average birth kick velocity of the pulsar. 

The production of pairs occurs through a fascinating process: the induced electric field due to rapid rotation extracts electrons from the star's surface, and these electrons find themselves in intense magnetic fields which lead to curvature radiation. In turn the photons produced in this way produce pairs and the process continues so as to form an electro-magnetic cascade which generates all the pairs we are interested in, though the maximum energy at this point is of order few GeV. In a young pulsar, the relativistic wind carrying these pairs away from the star eventually encounters the ejecta of the supernova so that a reverse shock is generated. This shock, referred to as the termination shock is the very site where the pairs are accelerated to very high energies. The PWNe that we observe are due to the radiation emitted by these accelerated pairs downstream of the termination shock. The luminosity of the PWNe is typically of the same order of the spin down luminosity of the central pulsar. As long as the pairs are in this complex environment, it is difficult to envision an obvious contribution to the cosmic ray positrons, as recognized by previous publications on this topic \cite{hooper}, both because of radiative and adiabatic energy losses. 

On the other hand, we have now clear evidence that pulsars remain inside the parent remnant only for $10^{4}-10^{5}$ years. After this time, the pulsar actually escapes the region limited by the blast wave of the supernova and moves with its initial kick velocity in the ISM. The typical velocity of $\sim 500$ km/s leads to supersonic motion, so that the relativistic wind that still emanates from the pulsar generates a bow shock and a nebula inside it. In a way similar to what happens when the pulsar is inside a remnant, a termination shock is also formed as a reverse shock and particle acceleration takes place around it. Although the acceleration process is all but known, on purely observational grounds we know that the spectrum of the accelerated pairs is a broken power law, with a low energy part (up to $\sim 100-1000$ GeV) as flat as $E^{-1}-E^{-1.8}$. The open structure of bow shock nebulae appears to be an excellent avenue for particle escape from the nebula. 

The energy still available in the form of pulsar rotation after the escape from the remnant is of order $\sim 10^{49}$ erg if the pulsar spins down as a magnetic dipole (braking index $n=3$) or of order $\sim 10^{47}$ in the case of braking index $n=2.5$, similar to the one measured in the Crab pulsar. 

The flux of electrons and positrons produced in the parent SNR (through acceleration at the shock) and at the pulsar wind termination shock after the pulsar escapes the remnant has been calculated by taking into account the random nature of supernova events (in space and time). The results are extremely interesting and can be summarized as follows:
\begin{itemize}
\item
For the case of dipole spin down one needs $\sim 1\%$ of the energy available in the pulsar rotation after escape from the remnant in order to explain at the same time the electrons (plus positrons) spectrum measured by Fermi and the positron ratio measured by PAMELA. The efficiency becomes $\sim 30-50\%$ for the case with braking index $n=2.5$. We want to point out that an efficiency of order unity is in a way more natural for the bow shock phase, since the pairs with energy below $\sim 1$ TeV or so can only end up in the ISM as cosmic ray particles. It would therefore be rather problematic to explain the fate of pairs in the dipole case, where the required efficiency is only $\sim 1\%$. On the other hand one should keep in mind that a dipole-like spin down is not observed in any of the pulsars for which the braking index is known. 

\item
As a consequence of the previous point, the contribution of pulsars to the positron flux at the Earth is basically unavoidable since the production rate is guaranteed (within the uncertainties associated with the spin down rate) and the efficiencies required to fit both the electron spectrum and the positron ratio are less than unity. In the dipole case, when the $\xi_{\pm}\ll 1$ one would even have the problem of explaining where the rest of the positrons are. 

\item
The fit to the total electron spectrum is made possible by the combination of the contribution of electrons from acceleration at the forward shock of supernovae (with a spectrum $E^{-2.4}$) and the contribution of electrons and positrons from pulsars, with flat injection spectrum. Both contributions are heavily affected by the local distribution of supernovae, especially at high energies, where energy losses limit the distance to the sources that contribute to the flux at the Earth. 

\item
The positron fraction observed by PAMELA is easily fit in all the configurations that we investigated, with the same efficiencies required to fit the total electron spectrum. The pulsar wind scenario also explains in a natural way the absence of antiprotons in excess of those produced in inelastic cosmic ray interactions in the Galaxy. 

\end{itemize}

\begin{acknowledgement}
This work was partially supported by ASI through contract ASI-INAF I/088/06/0.
\end{acknowledgement}

%%%%%%%%%%%%%%%%%%%%%%%% referenc.tex %%%%%%%%%%%%%%%%%%%%%%%%%%%%%%
% sample references
% %
% Use this file as a template for your own input.
%
%%%%%%%%%%%%%%%%%%%%%%%% Springer-Verlag %%%%%%%%%%%%%%%%%%%%%%%%%%
%
% BibTeX users please use
% \bibliographystyle{}
% \bibliography{}
%

\end{document}